  \documentclass{gretsi}
\usepackage{amssymb,amsmath,amsthm, mathtools}
\renewcommand\[{\begin{equation}}
\renewcommand\]{\end{equation}}

 \usepackage[latin1]{inputenc}   
 \usepackage[english,french]{babel}   
  \usepackage{times}			
 
 \usepackage[T1]{fontenc}

\usepackage{graphicx}
\usepackage{float}

\usepackage[T1]{tipa}

\usepackage{lipsum}
\newcommand\blfootnote[1]{%
  \begingroup
  \renewcommand\thefootnote{}\footnote{#1}%
  \addtocounter{footnote}{-1}%
  \endgroup
}

\titre{Transform\'{e}e en scattering sur la spirale temps-chroma-octave}

   \auteur{\coord{Vincent}{Lostanlen}{}, \coord{St{\'e}phane}{Mallat}{}}  
  \adresse{\affil{}{D{\'e}partement d'Informatique, {\'E}cole normale sup{\'e}rieure \\     %
     45 rue d'Ulm, Paris}}   %

\email{vincent.lostanlen@ens.fr}

\resumefrancais{
On introduit une reprÈsentation en scattering pour l'analyse et la classification des sons. Elle est localement invariante par translation, stable par dÈformation en temps et en frÈquence, et elle capture les structures harmoniques. Cette reprÈsentation en scattering peut s'interprÈter comme un rÈseau de neurones convolutif, calculÈ en cascadant une transformÈe en ondelettes dans le temps, et le long d'une spirale harmonique. Nous Ètudions son application pour l'analyse des dÈformations du modËle source-filtre.
}

\resumeanglais{
We introduce a scattering representation for the analysis and classification of sounds. It is locally translation-invariant, stable to deformations in time and frequency, and has the ability to capture harmonic structures. The scattering representation can be interpreted as a convolutional neural network which cascades a wavelet transform in time and along a harmonic spiral. We study its application for the analysis of the deformations of the source-filter model.
}

\begin{document}
\maketitle

\section{Introduction}

La variabilitÈ des signaux acoustiques naturels peut se modÈliser comme une action de dÈformation localisÈe en temps et en frÈquence.
Ainsi, la classification de sons repose essentiellement sur la construction de reprÈsentations qui demeurent stables ‡ ces dÈformations, tout en offrant une bonne discriminabilitÈ entre signaux de classes diffÈrentes.
En cascadant convolutions locales et non-linÈaritÈs, les reprÈsentations en rÈseaux de neurones parviennent ‡ combiner ces deux qualitÈs ; mais elles sont entiËrement adaptÈes aux donnÈes, et requiËrent par consÈquent une vaste base d'entraÓnement pour atteindre des performances satisfaisantes.

Dans cet article, nous proposons une reprÈsentation en cascade, dite transformÈe de scattering, dont l'architecture est similaire ‡ un rÈseau de neurones, mais sans besoin d'optimiser les unitÈs de convolution. On tire parti de la gÈomÈtrie naturelle des sons pour construire une description stable aux dÈformations et qui prÈserve l'information transitoire autant que possible.

\blfootnote{Ce travail est financ\'{e} par la bourse ERC InvariantClass 320959.
Le code source des exp\'{e}riences et figures est en libre acc\`{e}s
a l'adresse \texttt{www.github.com/lostanlen/scattering.m}.}
Un enjeu important de cette approche rÈside dans la prÈservation de la structure harmonique des partiels, y compris lorsque celle-ci est sujette ‡ des variations d'amplitude, de hauteur et de timbre.
Cette structure harmonique en peigne est trËs irrÈguliËre sur un axe log-frÈquentiel, et donc particuliËrement difficile ‡ caractÈriser dans un contexte polyphonique.

Pourtant, en enroulant l'axe log-frÈquentiel en une spirale, de sorte que les partiels sur des octaves consÈcutives se trouvent alignÈs, on fait apparaÓtre la rÈgularitÈ de l'enveloppe spectrale comme une dimension radiale. Une fois spÈcifiÈes les variables de temps, de chroma, et d'octave, le scattering en spirale consiste ‡ cascader trois dÈcompositions en ondelettes selon chacune de ces variables, puis ‡ appliquer le module complexe.
\begin{figure}
\label{spirale}
    \begin{center}
	\includegraphics[width=80mm]{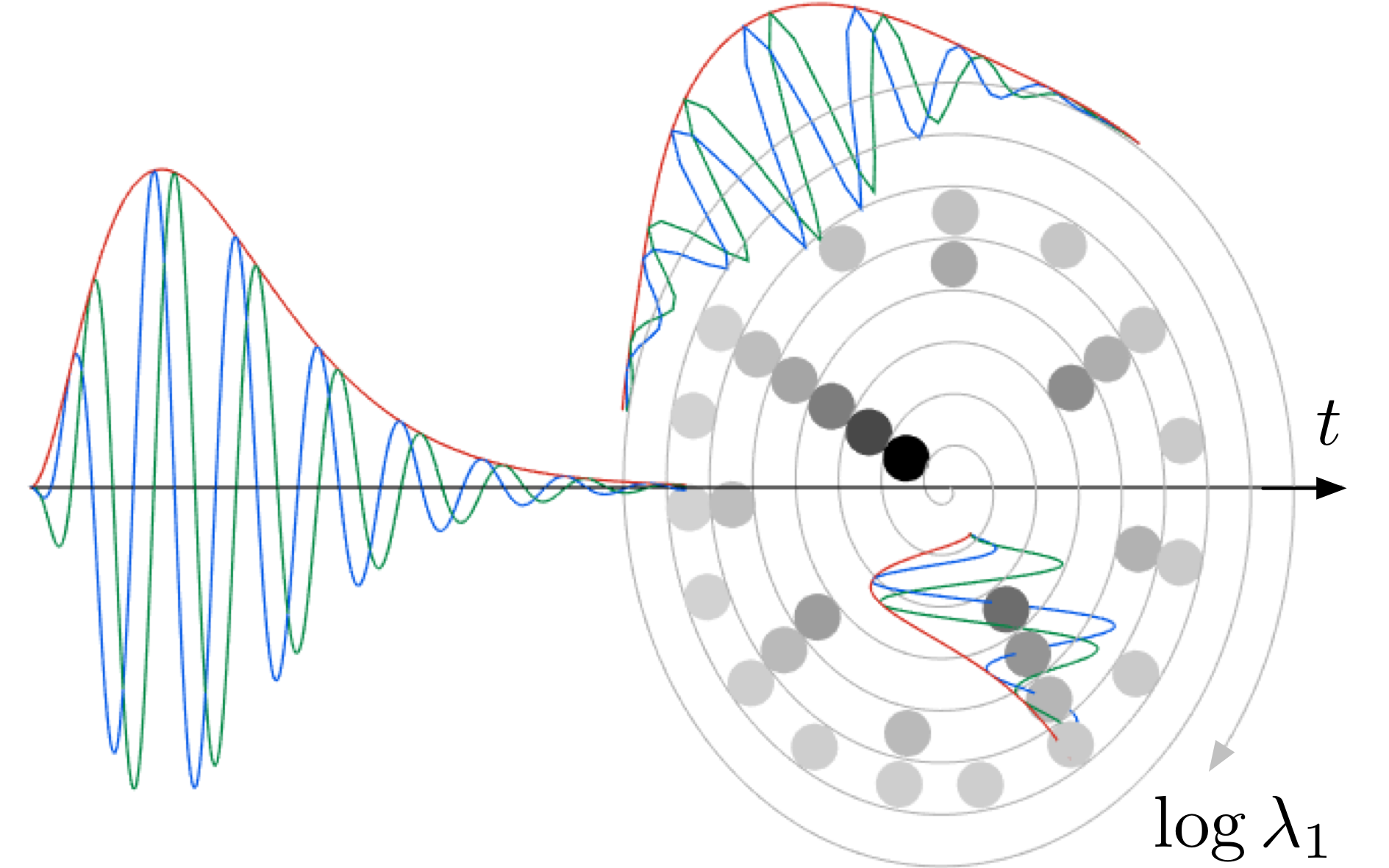}
    \end{center}
    \legende{L'ondelette en spirale est un produit d'ondelettes en temps, log-frÈquence, et octave. Les oscillations bleues et vertes reprÈsentent la partie rÈelle et la partie imaginaire. L'enveloppe rouge reprÈsente le module complexe. Les partiels d'un son harmonique, reprÈsentÈs en gris, suivent un motif d'alignement radial.}
    \label{fig:spiral-wavelets-3d}
\end{figure}

\vspace{-1mm}
\section{TransformÈes sur le scalogramme}

Dans cette section, on dÈfinit progressivement la transformÈe en scattering sur la spirale comme une extension de la transformÈe en scattering temporelle et de la transformÈe en scattering jointe temps-frÈquence. Les trois transformÈes partagent le mÍme formalisme.

\subsection{Scalogramme et scattering temporel}

On commence par construire une transform\'{e}e en ondelettes couvrant
les fr\'{e}quences audibles. Soit $\psi(t)$ un filtre passe-bande
\`{a} de fr\'{e}quence centrale r\'{e}duite $1$
et de largeur de bande $1/Q$. On dilate la transform\'{e}e de Fourier
$\hat{\psi}(\omega)$ de
$\psi(t)$ par des facteurs de r\'{e}solution
$\lambda_1 = 2^{j_1 + \chi_1}$ o˘ $j_1 \in \mathbb{Z}$ et $\chi_1 \in \{1 \ldots Q\}$ : 
\begin{equation}
\widehat{\psi}_{\lambda_{1}}(\omega) = \widehat{\psi}({\lambda_1}^{-1} \omega),
\text{ soit }
\psi_{\lambda_{1}}(t)=\lambda_{1}\psi(\lambda_{1}t).
\end{equation}
Chaque $\psi_{\lambda_{1}}(t)$ est un filtre passe-bande de
fr\'{e}quence centrale $\lambda_{1}$, de largeur de bande $\lambda_{1}/Q$
et de support temporel $2Q/\lambda_{1}$.
On construit donc un banc de filtres ‡ Q constant, capable de produire une reprÈsentation temps-frÈquence stable et parcimonieuse \cite{AM11, PPS+12}.
On choisit $Q=16$ dans les figures
de cet article.

Soit $\overset{t}{\ast}$ l'opÈrateur de convolution sur la variable temporelle $t$. On appelle scalogramme de $x(t)$ le module de la transformÈe en ondelettes $(x \ast \psi_{\lambda_{1}})$, indexÈ par le logarithme en base $2$ de la frÈquence acoustique $\lambda_1$ :
\[
x_{1}(t,\log_2\lambda_{1})=\vert x\ast\psi_{\lambda_{1}}\vert(t).
\]
La transformÈe ‡ Q constant (CQT) $S_1 x$ correspond ‡ un filtrage passe-bas de $x_1$ par une fenÍtre $\phi_T$ de support $T$:
\[
S_1 x(t,\log_2\lambda_1) = x_1 \overset{t}{\ast} \phi_T = \vert x \overset{t}{\ast} \psi_{\lambda_1} \vert \ast \phi_T.
\label{eq:S1-plain}
\]
$S_1 x$ est ainsi rendu invariant ‡ toute translation infÈrieure ‡ $T$. Toutefois, lors de ce filtrage passe-bas, les modulations d'amplitude dans $x_1$ de frÈquence supÈrieure ‡ $1/T$ sont dÈtruites. Afin de les restaurer, AndÈn et Mallat \cite{AM11} ont introduit la transformÈe de scattering comme le scalogramme du scalogramme:
\[
x_2(t,\log_2 \lambda_1,\log_2 \lambda_2) = \vert x_1 \overset{t}{\ast} \psi_{\lambda_2} \vert = \left \vert \vert x \overset{t}{\ast} \psi_{\lambda_1 } \vert \overset{t}{\ast} \psi_{\lambda_2} \right \vert.
\label{eq:x2-plain}
\]
Les ondelettes $\psi_{\lambda_2}(t)$ ont un facteur de qualitÈ Ègal ‡ $1$, mais nous choisissons de conserver la notation $\psi$ par souci de simplicitÈ. Chaque ondelette $\psi_{\lambda_2}(t)$ a pour frÈquence centrale $\lambda_2$ et pour support temporel $2 / \lambda_2$.
Comme dans l'Èquation (\ref{eq:S1-plain}), le filtrage de $x_2$ par $\phi_T(t)$ crÈe une reprÈsentation $S_2 x$ invariante ‡ la translation jusqu'‡ $T$:
\[
S_2 x(t,\log_2 \lambda_1,\log_2 \lambda_2) =
\left \vert \vert x \overset{t}{\ast} \psi_{\lambda_1 } \vert \overset{t}{\ast} \psi_{\lambda_2} \right \vert \overset{t}{\ast} \phi_T.
\]

\subsection{TransformÈe jointe temps-Èchelle}

La transformÈe de scattering dÈfinie ‡ l'Èquation (\ref{eq:x2-plain}) dÈcompose chaque bande de frÈquence $\lambda_1$ indÈpendamment, et ne peut donc pas capturer la cohÈrence de structures sonores temps-frÈquence, telles que les variations de hauteur.
Pour y remÈdier, AndÈn \cite{And14} a redÈfini les ondelettes $\psi_{\lambda_2}$'s comme des fonctions du temps et de la log-frÈquence, indexÈes par la paire $\lambda_{2}=(\alpha,\beta)$, o\`u $\alpha$ est une frÈquence de modulation en Hertz et $\beta$ est une frÈquence sur les dÈplacements en log-frÈquence :
\[
\psi_{\lambda_{2}}(t,\log_2\lambda_{1})=\psi_{\alpha}(t)\times
\psi_{\beta}(\log_2\lambda_{1}).
\]
La variable $\beta$ est mesurÈe en cycles par octave ; elle peut prendre des valeurs positives ou nÈgatives, ce qui permet de reprÈsenter des changements de hauteur montants ou descendants.
Le support temporel de $\psi_{\lambda_2}$ est maintenant $2/\alpha$, tandis que son support log-frÈquentiel est $2/\beta$.
On note $\overset{\chi_1}{\ast}$ les convolutions selon l'axe log-frÈquentiel. La transformÈe en scattering est Ètendu au cadre \og{}joint\fg{} temps-Èchelle en remplaÁant $\psi_{\lambda_2}$ par $(\psi_{\alpha}\times\psi_{\beta})$ dans l'Èquation (\ref{eq:x2-plain}):
\[
x_{2}(t,\log_2 \lambda_{1},\log_2 \lambda_{2})=\vert x_{1}\ast\psi_{\lambda_{2}}\vert  = \vert x_1 \overset{t}{\ast} \psi_{\alpha} \overset{\chi_1}{\ast} \psi_{\beta} \vert.
\]

Le modËle joint temps-frÈquence correspond \`a la transform\'ee \og{}corticale\fg{} introduite par
Shamma \cite{PPS+12} afin de formaliser ses dÈcouvertes en neurologie de l'audition.

\subsection{TransformÈe sur la spirale}

\begin{figure}
    \begin{center}
	\includegraphics[height=7cm]{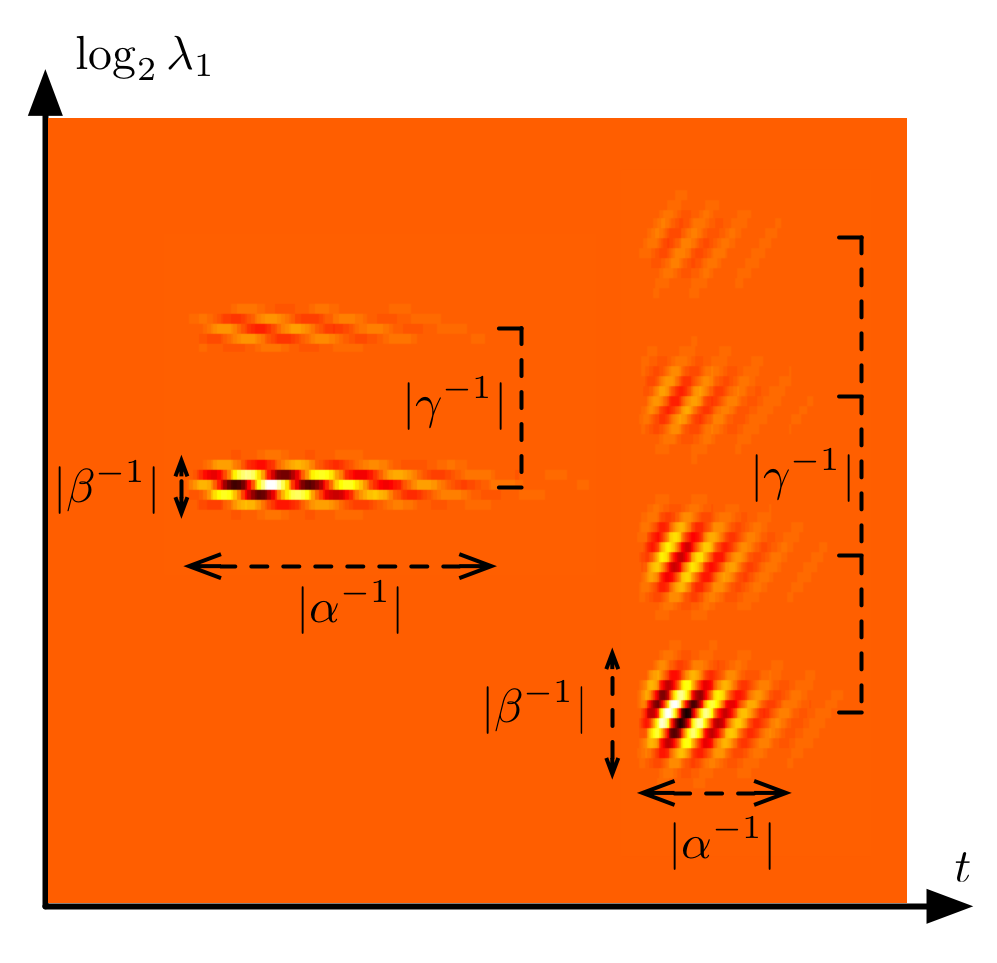}
    \end{center}
    \legende{Deux ondelettes en spirale $\psi_{\lambda_{2}}$ \'{e}tal\'{e}es sur le plan temps-fr\'{e}quence, pr\'{e}sentant des $\lambda_{2}=(\alpha,\beta,\gamma)$ diff\'{e}rents et une localisation diff\'{e}rente sur le scalogramme. \`{A} gauche : $\alpha^{-1}=120\,\mathrm{ms}$, $\beta^{-1}=-0.25\,\mathrm{octave}$, $\gamma^{-1}=+2\ \mathrm{octaves}$. \`{A} droite : $\alpha^{-1}=60\,\mathrm{ms}$, $\beta^{-1}=+0.5\,\mathrm{octave}$, $\gamma^{-1}=-4\,\mathrm{octaves}$. On a affich\'{e} la partie r\'{e}elle des coefficients. Le noir correspond \`{a} des coefficients positifs et le blanc \`{a} des coefficients n\'{e}gatifs.}
    \label{fig:spiral-wavelets-tfplane}
\end{figure}

La transformÈe jointe temps-Èchelle dÈcrit la variabilitÈ temporelle de hauteur sans recourir ‡ une segmentation prÈalable. Cependant, elle est agnostique ‡ la structure harmonique des sons voisÈs. L'Èvolution de cette structure recËle de l'information sur les formants en parole, ou sur les attaques instrumentales en musique par exemple. On peut la mesurer en comparant des partiels voisins sur des Èchelles en log-frÈquence allant de une ‡ quatre octaves, et ce ‡ chroma fixÈ. Nous proposons donc d'Ètendre la transformÈe jointe temps-frÈquence afin d'incorporer les dÈplacements sur les octaves en conjonction avec les dÈplacements sur les log-frÈquences voisines.

Conceptuellement, cela revient ‡ enrouler la variable de log-frÈquence $\log_2 \lambda _1$ selon la spirale des hauteurs (voir figure \ref{fig:spiral-wavelets-3d}) : on rÈvËle ainsi la variable radiale d'octave $j_1$ et la variable angulaire de chroma $\chi_1$.
En suivant le mÍme procÈdÈ que dans les deux transformÈes dÈfinies auparavant, on commence par dÈfinir une ondelette $\psi_{\lambda_2}$ comme un produit sÈparable d'ondelettes sur chacune des variables ‡ transformer.
Dans cet article, on a choisi une ondelette gammatone (profil asymÈtrique) selon le temps, une ondelette de Morlet (profil symÈtrique) selon les chromas et une ondelette gammatone selon les octaves.
\[
\psi_{\lambda_{2}}(t,\log\lambda_{1},j_1)=\psi_{\alpha}(t)\,\psi_{\beta}(\log\lambda_{1})\,\psi_{\gamma}(j_1).
\]
La figure \ref{fig:spiral-wavelets-tfplane} illustre la structure gÈomÈtrique de l'ondelette en spirale $\psi_{\lambda_{2}}$ dans le plan $(t,\log_2 \lambda_1)$, pour diffÈrentes valeurs de $\lambda_2 = (\alpha,\beta,\gamma)$.
Nous dÈfinissons la transformÈe en spirale comme une convolution sÈparable entre le scalogramme et $\psi_{\lambda_2}$, selon les trois variables de temps $t$, log-frÈquence $\log_2 \lambda_1$, et octave $j_1 = \lfloor \log_2 \lambda_1 \rfloor$ (partie entiËre) :
\[
x_{2}(t,\log\lambda_{1},\log\lambda_{2})=\vert x_{1}\ast\psi_{\lambda_{2}}
(t,\log \lambda_1,\lfloor \log \lambda_1 \rfloor)\vert .
\]
Il se trouve que l'idÈe consistant ‡ enrouler les hauteurs en spirale est bien connue en thÈorie de la musique, ne serait-ce que par la circularitÈ des noms de notes. Elle a notamment ÈtÈ ÈtudiÈe par Shepard et Risset pour construire des paradoxes de hauteurs \cite{Ris78} et a ÈtÈ validÈe par des imageries fonctionnelles du cortex auditif \cite{WUP+99}.

\section{DÈformations du modËle source-filtre}

Un modËle de production sonore classique consiste en la convolution d'un signal de source glottique $e(t)$ avec un filtre de conduit vocal $h(t)$. Dans cette section, on introduit une variabilitÈ de hauteur et d'enveloppe spectrale par des dÈformations temporelles de $e$ et $h$. On montre comment les propriÈtÈs d'harmonicitÈ de $e(t)$ et de rÈgularitÈ spectrale de $h(t)$, ÈnoncÈes ‡ l'Èquation (\ref{eq:properties}), permettent de sÈparer et linÈariser ces deux vitesses de dÈformation, sans Ètape de dÈtection prÈalable.

\subsection{RÈsultat principal}
Soit $\sum_{n}\delta(t-2 \pi n)$
un signal harmonique \og source \fg{} et soit $t\mapsto\theta(t)$ un diff\'{e}omorphisme
du temps ; on d\'{e}finit $e_{\theta}(t)=(e\circ\theta)(t)$ la source d\'{e}form\'{e}e.
De m\^{e}me, on compose un \og filtre \fg{} $h(t)$ et un diff\'{e}omorphisme
$t\mapsto\eta(t)$ pour d\'{e}finir $h_{\eta}(t)=(h\circ\eta)(t)$. Le
mod\`{e}le source-filtre d\'{e}form\'{e} est le signal
\[
x_{\theta,\eta}(t)=(e_{\theta}\ast h_{\eta})(t).
\]

La dÈrivÈe $\dot{\theta}(t)$ de $\theta(t)$ induit un changement de hauteur, tandis que $\dot{\eta}(t)$ provoque une dilatation locale de l'enveloppe spectrale $\vert \widehat{h}(\omega)\vert$. Nous allons montrer que, pour $\dot{\theta}(t)$ et $\dot{\eta}$ suffisamment rÈguliers sur le support des ondelettes de premier ordre $\psi_{\lambda_1}$, les maxima locaux de $x_2$ sont rassemblÈs sur un plan de l'espace $(\alpha,\beta,\gamma)$ des coefficients de scattering en spirale. Ce plan satisfait l'Èquation cartÈsienne
\[
\alpha + \frac{\ddot{\theta}(t)}{\dot{\theta}(t)} + \frac{\ddot{\eta}(t)}{\dot{\eta}(t)} \gamma = 0.
\label{eq:cartesian-equation}
\]
Dans un contexte polyphonique, ce rÈsultat signifie que des sons se chevauchant en temps et en frÈquence pourraient Ítre distinguÈs par leurs vitesses respectives de source et de filtre.
Les deux propriÈtÈs essentielles qui le sous-tendent sont l'harmonicitÈ de $\hat{e}(\omega)$ et la rÈgularitÈ spectrale de $\hat{h}(\omega)$.

\subsection{Factorisation du scalogramme}
On s'intÈresse au comportement du modËle autour du $p^\text{iËme}$ partiel : soient $t$ et $\lambda_1$ tels que $\lambda_1$ est proche de $p \dot{\theta}(t)$. Afin de pouvoir linÈariser $\theta(t)$ et $\nu(t)$ sur le support de $\psi_{\lambda_1}$, on travaille sous les hypothËses suivantes :
\begin{enumerate}
\item $\Vert\ddot{\eta}/\dot{\eta}\Vert_{\infty}\ll\lambda_{1}/Q$
(filtre lentement variable),
\item $\Vert\mathrm{d}(\log \vert \hat{h}\vert)/ \mathrm{d}\omega\Vert_{\infty} \times \Vert1/\dot{\eta}\Vert_{\infty}\ll Q/\lambda_{1}$
(rÈgularitÈ spectrale),
\item $\Vert\ddot{\theta}/\dot{\theta}\Vert_{\infty}\ll\lambda_{1}/Q$
(source lentement variable) et
\item $p<Q/2$ (partiel de rang faible).
\end{enumerate}
Les ÈgalitÈs (12) ‡ (14) sont des approximations de Taylor valables uniquement dans ce contexte.

Avec (a), on peut nÈgliger la contribution des partiels $p^{\prime} \neq p$ dans le scalogramme de $e_\theta$. Avec (b), la localisation temporelle de l'ondelette $\psi_{\lambda_1}(t)$ permet de remplacer l'action du diffÈomorphisme $\theta(t)$ est remplacÈe par une homothÈtie d'un facteur $\dot{\theta}(t)$ :
\[
\vert e_\theta \overset{t}{\ast} \psi_{\lambda_1} \vert (t) = \vert \widehat{\psi}_{\lambda_1}(p \dot{\theta}(t)) \vert.
\]
De mÍme, avec (c), l'action de $\eta(t)$ est remplacÈe par une homothÈtie d'un facteur $\dot{\eta}(t)$. Par ailleurs, avec (d), la localisation frÈquentielle de cette mÍme ondelette permet de remplacer l'enveloppe spectrale $\hat{h}(\omega)$ par une constante autour de la frÈquence $\lambda_1 / \dot{\eta}(t)$ :
\[
\vert h_\eta \overset{t}{\ast} \psi_{\lambda_1} \vert (t) = \hat{h}\left(\frac{\lambda_1}{\dot{\eta}(t)}\right) \times \psi_{\lambda_1}\left(\frac{\nu(t)}{\dot{\nu}(t)} \right).
\]
En menant ces deux linÈarisations conjointement, on aboutit ‡
\[
\vert x_{\theta,\eta} \overset{t}{\ast} \psi_{\lambda_1} \vert (t) =
\vert \widehat{\psi}_{\lambda_1}(p \dot{\theta}(t)) \vert
\times
\hat{h}\left(\frac{\lambda_1}{\dot{\eta}(t)}\right).
\]

\subsection{HarmonicitÈ et rÈgularitÈ spectrale}
Les ondelettes $\psi_\beta$ et $\psi_\gamma$ sont conÁues pour Ítre orthogonales aux fonctions affines. Or l'harmonicitÈ de $e_{\theta(t)}$ implique que son scalogramme est une constante le long de la variable d'octave $j_1$, et ce pour tout diffÈomorphisme $\theta(t)$. De plus, la rÈgularitÈ spectrale de $h_{\eta}(t)$ implique que son scalogramme est quasi linÈaire le long de la variable de chroma $\chi_1$. Ces deux propriÈtÈs s'Ècrivent
\begin{equation}
\Big\vert \vert e_\theta \overset{t}{\ast} \psi_{\lambda_1} \vert \overset{j_1}{\ast} \psi_{\gamma} \Big\vert \approx 0
\quad
\text{ et }
\quad
\Big\vert \vert e_\theta \overset{t}{\ast} \psi_{\lambda_1} \vert \overset{\chi_1}{\ast} \psi_{\beta} \Big \vert \approx 0.
\label{eq:properties}
\end{equation}
La dÈfinition du scattering en spirale se factorise alors en
\begin{eqnarray}
x_{\theta,\nu}\overset{t,\chi_{1},j_{1}}{\ast}\psi_{\lambda_{2}}  \mkern-72mu \\
 = & \left(\Big(\vert e_{\theta}\overset{t}{\ast}\psi_{\lambda_{1}}\vert\overset{\chi_{1}}{\ast}\psi_{\beta}\Big)\times \Big(\vert h_{\eta}\overset{t}{\ast}\psi_{\lambda_{1}}\vert\overset{j_{1}}{\ast}\psi_{\gamma}\Big)\right)\overset{t}{\ast}\psi_{\alpha} \nonumber,
\end{eqnarray}
o˘ les opÈrateurs $\overset{t}{\ast}$, $\overset{\chi_1}{\ast}$, et $\overset{j_1}{\ast}$ dÈsignent des convolutions en temps, log-frÈquence, et octave respectivement.

\subsection{Extraction de frÈquences instantanÈes}
Pour terminer, on constate que la phase du scalogramme de la source $\vert e_\theta \overset{t}{\ast} \psi_{\lambda_1} \vert \overset{\chi_1}{\ast} \psi_{\beta}$ est $\beta \times (\log_2 {\lambda_1} - \log_2 (p \dot{\theta}(t))$. En dÈrivant cette quantitÈ ‡ $\log_2 \lambda_1$ fixÈ, on trouve une frÈquence instantanÈe Ègale ‡ $-\beta \ddot{\theta}(t)/\dot{\theta}(t)$. De mÍme, la frÈquence instantanÈe du scalogramme du filtre aprËs convolution selon les octaves est $- \gamma \ddot{\eta}(t) / \dot{\eta}(t)$. En supposant que
\[
\alpha \geq \left\vert \dfrac{\ddot{\theta}(t)}{\dot{\theta}(t)} \beta \right\vert
\quad
\text{et}
\quad
\alpha \geq \left\vert \dfrac{\ddot{\eta}(t)}{\dot{\eta}(t)} \gamma \right\vert,
\]
les enveloppes de ces deux convolutions sont approximativement constantes sur le support de $\psi_{\alpha}(t)$. On conclut avec la formule approchÈe suivante pour les coefficients de scattering en spirale du modËle source-filtre dÈformÈ :
\begin{eqnarray}
\mkern-60mu
x_2(t,\log \lambda_1, \log \lambda_2) = 
\Big\vert \vert e_\theta \overset{t}{\ast} \psi_{\lambda_1} \vert \overset{\chi_1}{\ast} \psi_{\beta} \Big\vert
\times
\Big\vert \vert h_\nu \overset{t}{\ast} \psi_{\lambda_1} \vert \overset{j_1}{\ast} \psi_{\gamma} \Big\vert \nonumber \mkern-310mu \\
& \times \left \vert \widehat{\psi}_{\alpha} \left( -\dfrac{\ddot{\theta}(t)}{\dot{\theta}(t)} \beta - \dfrac{\ddot{\nu}(t)}{\dot{\nu}(t)} \gamma \right) \right \vert.
\label{eq:x2-sourcefilter}
\end{eqnarray}
Le spectre $\vert\widehat{\psi}_{\alpha}(\omega)\vert$ de $\psi_{\alpha}(t)$ est une bosse centrÈe en $\alpha$. L'Èquation (\ref{eq:cartesian-equation}) est une consÈquence immÈdiate de la formule ci-dessus. Ce rÈsultat reste vrai aprËs filtrage passe-bas par $\phi_T$ ‡ condition que les vitesses $\ddot{\theta}/\dot{\theta}(t)$ et $\ddot{\eta}(t)/\dot{\eta}(t)$ aient des variations relatives lentes devant $T$ :
\begin{equation}
\left \vert \frac{\dddot{\theta}(t)}{\ddot{\theta}(t)} - \frac{\ddot{\theta}(t)}{\dot{\theta}(t)} \right \vert \ll T^{-1}
\quad \text{et} \quad
\left \vert \frac{\dddot{\nu}(t)}{\ddot{\nu}(t)} - \frac{\ddot{\nu}(t)}{\dot{\nu}(t)} \right \vert \ll T^{-1}.
\end{equation}

\subsection{Illustration numÈrique}
\begin{figure}[h!]
    \begin{center}
	\includegraphics{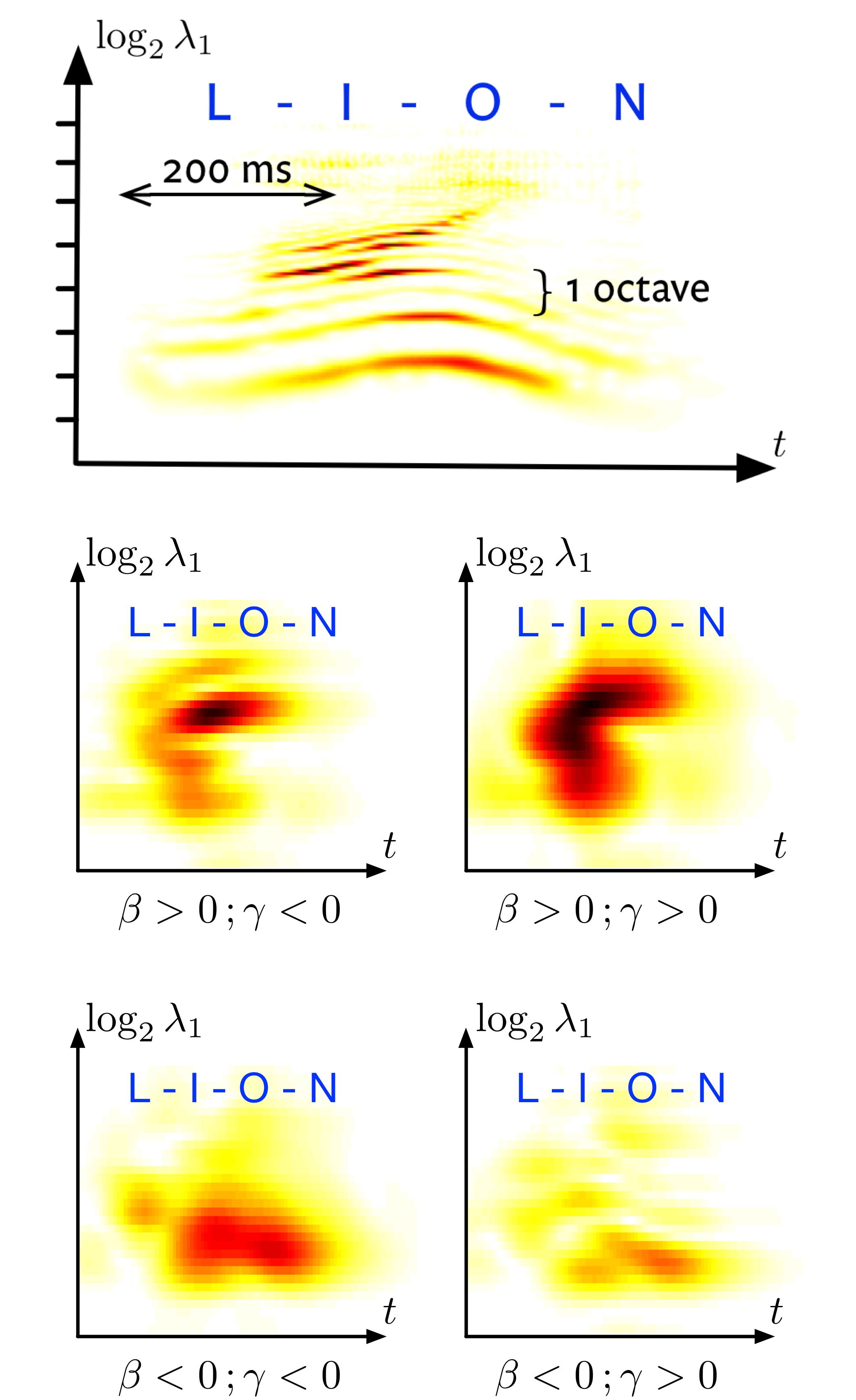}
    \end{center}
    \legende{En haut, un scalogramme $x_{1}(t,\log_2 \lambda_{1})$ du mot anglais \emph{lion}
    (prononcÈ /\textprimstress la\textsci \textschwa n/ ).
    En bas, coefficients de scattering de $x_{2}(t,\log_2 \lambda_{1},\log_2 \lambda_{2})$
en fonction du temps $t$ et de la log-fr\'{e}quence $\log_{2}\lambda_{1}$,
pour $\lambda_{2}=(\alpha,\beta,\gamma)$ fix\'{e} avec $\alpha^{-1}=120\,\mathrm{ms}$,
$\beta^{-1}=\pm1\,\mathrm{octave}$, $\gamma^{-1}=\pm4\,\mathrm{octaves}$.
La clart\'{e} est inversement proportionelle \`{a} l'amplitude des
coefficients.}
\label{fig:lion}
\end{figure}

La figure \ref{fig:lion} illustre le comportement temps-frÈquence de certains coefficients de scattering en spirale pour le mot anglais $\emph{lion}$, prononcÈ /\textprimstress la\textsci \textschwa n/ .
On constate que la syllabe /\textprimstress la\textsci / active en
particulier les coefficients tels que $\beta>0$, $\gamma>0$ (hauteur montante,
timbre montant) tandis que /\textsci \textschwa n/ active les coefficients
tels que $\beta<0$, $\gamma<0$ (hauteur descendante, timbre descendant). Ces signes sont corrÈlÈs avec
les sens de dÈformations du modËle source-filtre : $\ddot{\theta}(t)<0$ et $\ddot{\eta}(t)<0$ pour la syllabe /\textprimstress la\textsci /, $\ddot{\theta}(t)>0$ et $\ddot{\eta}(t)>0$ pour la syllabe  /\textsci \textschwa n/.

\section{Conclusion}
Le modËle en spirale prÈsentÈ ici est bien connu en musique et en psychologie expÈrimentale. Cependant, les mÈthodes existantes en traitement du signal ne tirent pas avantage de sa richesse : elles reprÈsentent la hauteur sur une ligne (MFCC) ou sur un cercle (vecteurs de chroma). Dans cet article, on a montrÈ comment la transformÈe de scattering sur la spirale caractÈrise les transitoires des sons harmoniques.

\end{document}